# Privacy, freedom of expression, and the right to be forgotten in Europe

Stefan Kulk & Frederik Zuiderveen Borgesius[1]

S.Kulk [at] uu.nl & F.J.ZuiderveenBorgesius [at] uva.nl

**Abstract** – In this chapter we discuss the relation between privacy and freedom of expression in Europe. In principle, the two rights have equal weight in Europe – which right prevails depends on the circumstances of a case. We use the *Google Spain* judgment of the Court of Justice of the European Union, sometimes called the 'right to be forgotten' judgment, to illustrate the difficulties when balancing the two rights. The court decided in *Google Spain* that people have, under certain conditions, the right to have search results for their name delisted. We discuss how Google and Data Protection Authorities deal with such delisting requests in practice. Delisting requests illustrate that balancing privacy and freedom of expression interests will always remain difficult.

---

[1] As in their earlier publications, both authors equally contributed to this chapter. Stefan Kulk is a researcher at the Utrecht Centre for Accountability and Liability Law (UCALL), and the Centre for Intellectual Property Law (CIER) of Utrecht University. Dr. Frederik Zuiderveen Borgesius is a researcher at the Institute for Information Law (IViR) of the University of Amsterdam, and at the Research Group on Law, Science, Technology & Society (LSTS), of the Vrije Universiteit Brussels. We would like to thank David Erdos, Bojana Kostic, Marijn Sax, Nico van Eijk, Joris van Hoboken, and Kyu Ho Youm for their helpful comments. Any errors are our own.

## Table of Contents

## 1 Introduction

In this chapter, we discuss the relationship between privacy and freedom of expression in Europe. For readers who are not familiar with the complex legal order in Europe, we introduce the Council of Europe and its European Court of Human Rights (Section 2), and the European Union and its Court of Justice (Section 3). We discuss how those two courts deal with privacy and freedom of expression. We illustrate the tension between these two fundamental rights by looking at the judgment of the Court of Justice of the European Union in the *Google Spain* case, sometimes called the 'right to be forgotten' case (Section 4). The court decided in *Google Spain* that people have, under certain conditions, the right to have search results for their name delisted.

We then describe the development of the right to be forgotten, and discuss some open questions (section 5). Delisting requests illustrate that a case-by-case analysis is required when balancing the rights to privacy and freedom of expression (section 6). We can expect much more case law, that hopefully provides more guidance on how to strike the balance between these fundamental rights.

We consider how the two most important European courts deal with balancing privacy and freedom of expression.[2] We focus on the right to be forgotten, especially in its sense as right to have search results removed. The General Data Protection Regulation (GDPR),[3] applicable from 2018, also introduces a broader 'right to erasure'.[4] That GDPR provision is outside the scope of this chapter.[5]

## 2 The Council of Europe and its European Court of Human Rights

The Council of Europe was founded in 1949 just after the Second World War and is the most important human rights organisation in Europe. It is based in Strasbourg and now has 47 Member States.[6] In 1950, the Council of Europe adopted the European Convention on Human Rights.[7] This Convention lists human rights that Member States must guarantee. The Council of Europe also set up a court: the European Court of

---

[2] There are different traditions in Europe countries regarding the right balance between privacy and freedom of expression. See M Verpaux, *Freedom of Expression* (Europeans and their rights, Council of Europe 2010), p. 17-27; p. 197-199.

[3] European Parliament and Council Regulation (EU) 2016/679 of 27 April 2016 on the protection of natural persons with regard to the processing of personal data and on the free movement of such data, and repealing Directive 95/46/EC (General Data Protection Regulation) [2016] OJ L 119/1.

[4] See on the right to erasure ('to be forgotten') in the GDPR: JVJ Van Hoboken, 'The Proposed Right to be Forgotten Seen from the Perspective of Our Right to Remember. Freedom of Expression Safeguards in a Converging Information Environment (Prepared for the European Commission)' (May 2013); C Bartolini L and Siry, 'The right to be forgotten in the light of the consent of the data subject' (2016) 32(2) Computer Law & Security Review 218.

[5] The chapter builds on and borrows some sentences from our earlier work: S Kulk and FJ Zuiderveen Borgesius, 'Google Spain v. Gonzalez: Did the Court Forget about Freedom of Expression' (2014) 5(3) European Journal of Risk Regulation 389; S Kulk and FJ Zuiderveen Borgesius, 'Freedom of Expression and 'Right to Be Forgotten' Cases in the Netherlands after Google Spain' (2015) 1(2) European Data Protection Law Review 113; FJ Zuiderveen Borgesius and A Arnbak, 'New Data Security Requirements and the Proceduralization of Mass Surveillance Law after the European Data Retention Case' (October 23, 2015). Amsterdam Law School Research Paper No. 2015-41. <http://ssrn.com/abstract=2678860>; FJ Zuiderveen Borgesius, *Improving privacy protection in the area of behavioural targeting* (Kluwer Law International 2015).

[6] Council of Europe, 'Who We Are' <www.coe.int/en/web/about-us/who-we-are>, accessed 6 September 2017.

[7] European Convention on Human Rights [1950].

Human Rights, based in Strasbourg. That court decides on alleged violations of the rights in the European Convention on Human Rights.[8]

## 2.1 Privacy

The European Convention on Human Rights contains a right to privacy (Article 8)[9] and a right to freedom of expression (Article 10). Article 8 of the Convention protects the right to private life. (In this chapter, we use 'privacy' and 'private life' interchangeably.[10]) Article 8 in principle prohibits interferences with the right to privacy. Yet, paragraph 2 shows that this prohibition is not absolute. In many cases the right to privacy can be limited by other interests, such as public safety, or for the rights of others, such as freedom of expression. Article 8 reads as follows:

> 1. Everyone has the right to respect for his private and family life, his home and his correspondence.
>
> 2. There shall be no interference by a public authority with the exercise of this right except such as is in accordance with the law and is necessary in a democratic society in the interests of national security, public safety or the economic well-being of the country, for the prevention of disorder or crime, for the protection of health or morals, or for the protection of the rights and freedoms of others.

The European Court of Human Rights interprets the right to privacy generously, and refuses to define the ambit of the right.[11] The court 'does not consider it possible or necessary to attempt an exhaustive definition of the notion of private life.'[12] The court says it takes 'a pragmatic, common-sense approach rather than a formalistic or purely legal one.'[13] The court uses a 'dynamic and evolutive' interpretation of the Convention, and says 'the term "private life" must not be interpreted restrictively.'[14] In several cases, the court acknowledged that people also have a right to privacy when they are in public, such as in restaurants[15] or on the street.[16] The court's dynamic approach towards the interpretation of the Convention has been called the 'living instrument doctrine'.[17] The living instrument doctrine could be seen as the opposite of the US doctrine of originalism. The latter doctrine entails that the US Constitution is to be interpreted according to the original meaning that it had at the time of ratification.[18] In sum, the European Court of Human Rights gives extensive protection to the right to privacy interests, but the right is not absolute.

---

[8] Article 19 and 34 of the European Convention on Human Rights. In exceptional situations, the Grand Chamber of the Court decides on cases (Article 30 and 43 of the European Convention on Human Rights). Judgments by the Grand Chamber have more weight than judgments of other chambers of the Court.

[9] Article 8 of the European Convention on Human Rights: 'Right to respect for private and family life'.

[10] See on the difference between 'privacy' and 'private life': G González Fuster, *The Emergence of Personal Data Protection as a Fundamental Right of the EU* (Springer 2014) 82-84; 255.

[11] See generally on the Article 8 case law of the European Court of Human Rights: D Harris and others, *Law of the European Convention on Human Rights* (Oxford University Press, 2014) 522-591; R Ivana, *Protecting the right to respect for private and family life under the European Convention on Human Rights* (Council of Europe 2012); AW Heringa and L Zwaak, 'Right to respect for privacy' in P Van

## 2.2 Freedom of expression

The right to freedom of expression is protected in Article 10 of the European Convention on Human Rights. Paragraph 2 of Article 10 permits limitations on the right to freedom of expression, similar to paragraph 2 of Article 8. Hence, the right to freedom of expression may be limited 'for the protection of the reputation or rights of others',[19] including the right to privacy. Article 10 reads as follows:

---

Dijk and others (ed), *Theory and practice of the European Convention on Human Rights* (Intersentia 2006).

[12] See e.g. *Niemietz v Germany* App no 13710/88 (ECtHR 16 December 1992), para 29. The court consistently confirms this approach. See e.g. *Pretty v United Kingdom*, App no 2346/02 (ECtHR 29 April 2002), para 61; *Marper v United Kingdom* App no 30562/04 and 30566/04 (ECtHR 4 December 2008) para 66.

[13] *Botta v. Italy* App no 21439/93 (ECtHR 24 February 1998) para 27.

[14] *Christine Goodwin v United Kingdom*, App no 28957/95 (ECtHR 11 July 2002), para 74; Amann v. Switzerland, App no 27798/95 (ECtHR 16 February 2000), para 65.

[15] *Von Hannover v Germany (I)*, App no 59320/00 (ECtHR 24 September 2004).

[16] *Peck v the United kingdom*, App no 44647/98 (ECtHR 28 January 2003). See also N Moreham, 'Privacy in public places' (2006) 65(3) The Cambridge Law Journal 606.

[17] A Mowbray, 'The creativity of the European Court of Human Rights' (2005) 5(1) Human Rights Law Review 57. The court puts it as follows: 'That the Convention is a living instrument which must be interpreted in the light of present-day conditions is firmly rooted in the Court's case-law' (Matthews v. United Kingdom App no 24833/94 (ECtHR 18 February 1999), para 39). The court started the 'living instrument' approach in: *Tyrer v United Kingdom*, App no 5856/72 (ECtHR 25 April 1978) para 31.

[18] B Boyce, 'Originalism and the Fourteenth Amendment' (1998) 33 Wake Forest Law Review 909.

[19] Paragraph 2 of Article 10 of the European Convention on Human Rights.

> 1. Everyone has the right to freedom of expression. This right shall include freedom to hold opinions and to receive and impart information and ideas without interference by public authority and regardless of frontiers. This Article shall not prevent States from requiring the licensing of broadcasting, television or cinema enterprises.
>
> 2. The exercise of these freedoms, since it carries with it duties and responsibilities, may be subject to such formalities, conditions, restrictions or penalties as are prescribed by law and are necessary in a democratic society, in the interests of national security, territorial integrity or public safety, for the prevention of disorder or crime, for the protection of health or morals, for the protection of the reputation or rights of others, for preventing the disclosure of information received in confidence, or for maintaining the authority and impartiality of the judiciary.

Article 10 does not only protect the 'speaker' but also the 'listener', who has the right to *receive* information. As the European Court of Human Rights puts it, 'the public has a right to receive information of general interest.'[20] The court also notes that 'Article 10 applies not only to the content of information but also to the means of transmission or reception since any restriction imposed on the means necessarily interferes with the right to receive and impart information.'[21] Moreover, 'the internet plays an important role in enhancing the public's access to news and facilitating the sharing and dissemination of information generally (…).'[22] Or as the European Court of Human Rights highlighted in another case, 'user-generated expressive activity on the Internet provides an unprecedented platform for the exercise of freedom of expression.'[23]

---

[20] *Társaság a Szabadságjogokért v Hungary*, App no 37374/05 (ECtHR 14 April 2009), para. 26.
[21] *Autronic v Switzerland* App no 12726/87 (ECtHR 22 May 1990), para 47.
[22] *Fredrik Neij and Peter Sunde Kolmisoppi v Sweden* App no 40397/12 (ECtHR 19 February 2013), p 9.
[23] *Delfi v. Estonia* App no 64569/09 (ECtHR 16 june 2015), para. 110.

Privacy and freedom of expression have equal weight in the case law of the European Court of Human Rights: 'as a matter of principle these rights deserve equal respect.'[24] It depends on the circumstances in a particular case which right should prevail. The court has developed a large body of case law on balancing privacy and freedom of expression. The court takes a nuanced approach, taking all circumstances of a case into account.[25]

To balance privacy and freedom of expression, the European Court of Human Rights has developed a set of criteria. For instance, expression that advances public debate receives extra protection in the court's case law: an 'essential criterion is the contribution made by photos or articles in the press to a debate of general interest.'[26] And if a publication concerns a politician or a similar public figure, rather than an ordinary citizen, the European Court of Human Rights is more likely to rule that freedom of expression outweighs privacy.[27] The court summarises the main criteria as follows:

> 'where the right to freedom of expression is being balanced against the right to respect for private life, the relevant criteria in the balancing exercise include the following elements: contribution to a debate of general interest, how well known the person concerned is, the subject of the report, the prior conduct of the person concerned, the method of obtaining the information and its veracity, the content, form and consequences of the publication, and the severity of the

---

[24] *Axel Springer AG v Germany* App no 39954/08 (ECtHR 7 February 2012), para 87. See similarly: *Von Hannover v Germany* App nrs 40660/08 and 60641/08 (ECtHR 7 February 2012), para 100; *Węgrzynowski and Smolczewski v Poland* App no 33846/07 (ECtHR16 July 2013), para 56. See on this balancing approach: E Barendt, 'Balancing freedom of expression and privacy: the jurisprudence of the Strasbourg Court' (2009) 1(1) Journal of Media Law 49.

[25] M Oetheimer, *Freedom of Expression in Europe: Case-law Concerning Article 10 of the European Convention of Human Rights* (Council of Europe publishing 2007).

[26] *Axel Springer AG v Germany* App no 39954/08 (ECtHR 7 February 2012), para 90.

[27] *Axel Springer AG v Germany* App no 39954/08 (ECtHR 7 February 2012), para 91.

> sanction imposed [on the party invoking freedom of expression].'[28]

The Convention's provisions primarily protect people against their states. States should not interfere too much in people's lives – states thus have a negative obligation towards their citizens. States may only interfere with a person's right to freedom of expression if such an interference is proportionate and necessary in a democratic society. However, in the late 1960s the European Court of Human Rights started to recognize that the Convention could also imply positive obligations.[29] Hence, on some occasions, states must also take action to protect people against breaches of their human rights.[30] For instance, a state may have an obligation to protect journalists from violent attacks. If the state fails to meet that obligation, the state infringes the right to freedom of expression.[31]

People cannot bring a claim against other people before the European Court of Human Rights.[32] But people can complain to the court if their state does not adequately protect their rights against infringements by other people. The rights in the European Convention on Human Rights can also indirectly protect people against human rights violations by other non-state actors.

To illustrate how the European Convention on Human Rights balances privacy and freedom of expression, we briefly discuss the 2012 *Axel Springer* case. Springer publishes the mass-circulation German daily newspaper Bild. In 2004, Bild wrote about the arrest of a well-known German actor during the Munich beer festival (*Oktoberfest*) for possession of 0.23 gram of cocaine. The actor plays a Police Superintendent in a popular TV series. Bild put a picture of the actor on its front page with the text: 'Cocaine! Superintendent [name] caught at the Munich beer festival' (publication 1). In a later publication (2), Bild reported that the actor was convicted for cocaine possession, and was fined 18.000 euro.

---

[28] *Satakunnan Markkinapörssi Oy And Satamedia Oy v Finland* App no 931/13 (ECtHR 21 July 2015), para 83.

[29] J Akandji-Kombe, *Positive obligations under the European Convention on Human Rights. A guide to the implementation of the European Convention on Human Rights* (Council of Europe publishing 2007).

[30] See e.g. *Z v Finland* App no 22009/93 (ECtHR 25 February 1997), para. 36; *Mosley v. United Kingdom*, App no 48009/08 (ECtHR 10 May 2011), para 106. See generally: J Akandji-Kombe, 'Positive obligations under the European Convention on Human Rights' (2007)(7) Human rights handbooks; P De Hert, 'From the principle of accountability to system responsibility? Key concepts in data protection law and human rights law discussions.' International Data Protection Conference 2011 <www.vub.ac.be/LSTS/pub/Dehert/410.pdf> accessed 6 September 2017.

[31] See e.g. *Gundem v Turkey* App no 23144/93 (ECtHR 16 march 2000).

[32] Article 34 of the European Convention on Human Rights.

The actor went to court, claiming Bild invaded his privacy. German courts decided, in several instances, that Springer violated the actor's privacy, because there was no public interest in knowing about his offence. The German courts prohibited further publication of the Bild article, and ordered Springer to pay a 1000 Euro penalty. German courts gave similar judgments regarding publication 2.

Springer went to the European Court of Human Rights, and claimed that Germany violated its right to freedom of expression. Germany and Springer agreed that Springer's freedom of expression was interfered with, that the interference was prescribed by law, and that the aim of the interference was legitimate: protecting the reputation and privacy of the actor. But the parties disagreed on whether the interference with freedom of expression was 'necessary in a democratic society' (see Article 10(2) of the European Convention on Human Rights).

The European Court of Human Rights confirms that freedom of expression, as a general principle, is essential for democracy:

> Freedom of expression constitutes one of the essential foundations of a democratic society and one of the basic conditions for its progress and for each individual's self-fulfilment. Subject to paragraph 2 of Article 10, it is applicable not only to 'information' or 'ideas' that are favourably received or regarded as inoffensive or as a matter of indifference, but also to those that offend, shock or disturb. Such are the demands of pluralism, tolerance and broadmindedness without which there is no 'democratic society'. As set forth in Article 10, freedom of expression is subject to exceptions, which must,

> however, be construed strictly, and the need for any restrictions must be established convincingly.[33]

The European Court of Human Rights reaffirms that freedom of expression can be limited in view of the rights of others, such as privacy. The court agrees with the German courts' assessment that Springer's sole interest in writing about the actor was that he was a well-known actor who was arrested. However, the court emphasises that he was arrested in public, during the *Oktoberfest*. Moreover, the actor had revealed details about his private life in a number of interviews. Therefore, his legitimate expectation of privacy was reduced.[34] Furthermore, the publications had a sufficient factual basis.

According to the European Court of Human Rights, a balancing exercise was needed between the publisher's right to freedom of expression, and the actor's right to privacy. The court says 'there is nothing to suggest that such a balancing exercise was not undertaken' by Springer. As Springer had received the information about the actor from the police, it did not have strong grounds for believing that it should preserve his anonymity. Therefore, says the court, Springer did not act in bad faith. Additionally, Springer's publications did not 'reveal details about [the actor's] private life, but mainly concerned the circumstances of and events following his arrest'.[35] Nor did the publications contain disparaging expression or unsubstantiated allegation.[36]

Regarding the severity of the sanctions imposed on Springer, the court notes that 'although these were lenient, they were capable of having a chilling effect on the applicant company.'[37] The European Court of Human Rights concludes that Germany violated the right to freedom of expression of the publisher Springer.[38] In short,

---

[33] *Axel Springer AG v Germany* App no 39954/08 (ECtHR 7 February 2012), para 78. Internal citations omitted.
[34] ibid, para 101. Internal citations omitted.
[35] ibid, para 108.
[36] ibid, para 108.
[37] ibid, para 109.
[38] ibid, para 110-111.

Springer's freedom of expression right prevails, in this case, over the actor's privacy right.

In sum, the European Convention on Human Rights protects both privacy and freedom of expression. The Convention's privacy and freedom of expression rights can have a horizontal effect, which means that these rights are also relevant in disputes among citizens. The European Court of Human Rights says privacy and freedom of expression are equally important.

## 3 The European Union and its Court of Justice

The European Union has its origin in the European Coal and Steel Community, which was formed in 1951, and the European Economic Community and European Atomic Energy Community, which were formed in 1957. These communities and other forms cooperation developed into the European Union (EU), which was formally established in 1992 by the Maastricht Treaty. The EU has grown into an economic and political partnership between 28 (soon 27)[39] European countries. The 2007 Lisbon Treaty was the latest treaty to structurally reform the European Union.

The EU itself is not a party to the European Convention of Human Rights.[40] However, each of the EU member states is also a member of the Council of Europe, and must thus also adhere to the Convention on Human Rights.[41]

The Court of Justice of the European Union is one of the core EU institutions, and is based in Luxembourg.[42] National judges in the EU can, and in some cases must, ask the Court of Justice of the European Union for a preliminary judgment concerning the interpretation of EU law.[43] As noted, the European Union has its roots in economic cooperation. Until 1969 the Court of Justice of the European Union did not consider itself competent to rule on fundamental rights.[44] Nowadays, the Treaty on the European Union codifies the importance of fundamental rights.[45] Article 6(3) of the Treaty reads: 'Fundamental rights, as guaranteed by the European Convention for the Protection of Human Rights and Fundamental Freedoms and as they result from the constitutional traditions common to the Member States, shall constitute general principles of the Union's law.' Moreover, in 2000 the Charter of Fundamental Rights of the European Union was adopted. It lists the fundamental rights and freedoms recognized by the EU.[46] Since the Charter became a legally binding instrument in 2009,[47] the number of cases in which the Court of Justice of the European Union cited the Charter increased substantially.[48] Recently, the court has also given influential privacy and data protection-related judgments,[49] such as the *Google Spain* judgment (see Section 3.3 below).

---

[39] The United Kingdom is likely to leave the EU: the so-called Brexit.

[40] See also Opinion 2/13, ECLI:EU:C:2014:2454, in which the European Court of Justice advised against accession of the EU to the European Convention on Human Rights.

[41] Council of Europe, 'Our member States' <www.coe.int/en/web/about-us/our-member-states>, accessed 6 September 2017.

[42] Article 13(1) of the Treaty on European Union (consolidated version) ([2016] OJ C 202).

[43] Article 19(3)(b) of the Treaty on European Union (consolidated version).

## 3.1 Privacy and freedom of expression

The Charter of Fundamental Rights of the European Union contains a right to privacy and a right to freedom of expression that resemble the corresponding rights in the

---

[44] In 1969, the Court of Justice of the European Union said that 'fundamental human rights [are] enshrined in the general principles of Community law and protected by the Court' (Case C-29/69 *Stauder v Stadt Ulm* [1969] ECR 419, para 7). Also see: G González Fuster, *The Emergence of Personal Data Protection as a Fundamental Right of the EU* (Springer 2014) 164.

[45] Article 6(3) of the Treaty on the European Union (consolidated version).

[46] The phrases 'fundamental rights' and 'human rights' are roughly interchangeably. See on the slight difference: G González Fuster, *The Emergence of Personal Data Protection as a Fundamental Right of the EU* (Springer 2014) 164-166.

[47] Article 6(1) of the Treaty on European Union (consolidated version).

[48] G de Búrca, 'After the EU Charter of Fundamental Rights: the Court of Justice as a Human Rights Adjudicator?', (2013) 20(2) Maastricht Journal of European and Comparative Law 168.

[49] See e.g. joined cases C-293/12 and C-594/12 *Digital Rights Ireland and Seitlinger and others v Minister for Communications and others*, ECLI:EU:C:2014:238, invalidating the Data Retention Directive; and Case C-362/14 *Maximillian Schrems v Data Protection Commissioner,* ECLI:EU:C:2015:650, invalidating the Safe Harbor agreement. See generally: L *Laudati Summaries of EU court decisions relating to data protection 2000-2015* (OLAF European Anti-Fraud Office 2016) <https://ec.europa.eu/anti-fraud/sites/antifraud/files/caselaw_2001_2015_en.pdf> accessed 6 September 2017.

European Convention on Human Rights.[50] In addition to the right to privacy, the Charter contains a separate right to the protection of personal data.[51]

### 3.2 Data protection law

Article 8 of the Charter of Fundamental Rights of the European Union grants people the right to protection of personal data:

> 1. Everyone has the right to the protection of personal data concerning him or her.
>
> 2. Such data must be processed fairly for specified purposes and on the basis of the consent of the person concerned or some other legitimate basis laid down by law. Everyone has the right of access to data which has been collected concerning him or her, and the right to have it rectified.
>
> 3. Compliance with these rules shall be subject to control by an independent authority.

Since the 1990s, the EU has played an important role in the field of data protection law. Data protection law developed in response to the increasing amounts of personal information that were gathered by the state and large companies, typically using computers.[52] In the 1970s, several European countries adopted data protection laws. Some of those national data protection laws contained restrictions on the export of personal data. National lawmakers wanted to prevent that their citizen's data would be exported to countries without sufficient legal protection of personal data.[53]

[50] See Article 7 (privacy) and Article 11 (freedom of expression) in the Charter of Fundamental Rights of the European Union. See Article 52 in the Charter on the possible limitations on the exercise of the Charter's rights.

[51] The European Convention on Human Rights does not explicitly protect personal data. However, many cases that concern personal data are also covered by the Convention's right to privacy. See: P. de Hert and S. Gutwirth, 'Data Protection in the Case Law of Strasbourg and Luxemburg: Constitutionalisation in Action' in S. Gutwirth et al. (eds), *Reinventing data protection?* (Springer 2009).

[52] CJ Bennett, *Regulating privacy: data protection and public policy in Europe and the United States* (Cornell University Press 1992).

In 1981, the Council of Europe (not the European Union) adopted the first legally binding international instrument on data protection, the Data Protection Convention.[54] The Data Protection Convention requires signatories to enact data protection provisions in their national law.[55] The European Commission, which is a European Union institution, had called on Member States to ratify the Data Protection Convention in 1981. However, in 1990 only seven Member States had done so.[56] Moreover, the Data Protection Convention left possibilities for countries to raise barriers for personal data flows at the borders.[57] Many stakeholders feared that national authorities would stop the export of personal data to other European countries.

In 1990, the EU stepped in to harmonise data protection law in the European Union, and presented a proposal for a Data Protection Directive.[58] After five years of debate, the EU adopted 'Directive 95/46/EC of the European Parliament and of the Council of 24 October 1995 on the protection of individuals with regard to the processing of personal data and on the free movement of such data.'[59] This Data Protection Directive is probably the most influential data privacy text in the world.[60]

---

[53] ibid. See also FW Hondius, *Emerging Data Protection in Europe* (North-Holland Publishing Company 1975); V Mayer-Schönberger, 'Generational development of data protection in Europe' in PE Agre M and Rotenberg (eds), *Technology and Privacy: The new landscape* (MIT Press 1997).
[54] Convention for the Protection of Individuals with regard to Automatic Processing of Personal Data CETS No.: 108, 28 January 1981. The Convention is under revision: see <http://www.coe.int/en/web/data-protection/modernisation-convention108>.
[55] Article 4(1) of the Data Protection Convention.
[56] European Commission, Recommendation 81/679/EEC of 29 July 1981 relating to the Council of Europe Convention for the protection of individuals with regard to the automatic processing of personal data [1981] OJ L246/31; N Platten, 'Background to and History of the Directive' in D Bainbridge (ed), *EC Data Protection Directive* (Butterworth 1996) 17-18, 23.
[57] Article 12.3 of the Data Protection Convention allows states to derogate from the prohibition of interfering with cross border data flows, in brief because of the special nature of personal data, or to avoid circumvention of data protection law.
[58] PM Schwartz, *Managing Global Data Privacy: Cross-Border Information Flows in a Networked Environment* (Privacy Projects 2009) 11; ACM Nugter, *Transborder Flow of Personal Data within the EC. A Comparative Analysis of Princiepstat* (Kluwer Law International 1990).
[59] European Parliament and Council (EC) 95/46 on the Protection of Individuals with Regard to the Processing of Personal Data and on the Free Movement of Such Data [1995] OJ L281/31.
[60] See M Birnhack, 'The EU Data Protection Directive: An Engine of a Global Regime' Computer Law & Security Review (2008) 24(6) 508; M Birnhack, 'Reverse Engineering Informational Privacy Law' 15

The Data Protection Directive has two aims. First: to 'protect the fundamental rights and freedoms of natural persons, and in particular their right to privacy with respect to the processing of personal data.'[61] Second: to safeguard the free flow of personal data between EU member states. Under the Data Protection Directive, EU 'Member States shall neither restrict nor prohibit the free flow of personal data between Member States for reasons connected with the protection [of personal data].'[62]

EU data protection law grants rights to people whose data are being processed (data subjects),[63] and imposes obligations on parties that process personal data (data controllers).[64] The Data Protection Directive contains principles for fair data processing, comparable to the Fair Information Practice Principles.[65]

For instance, personal data must be processed lawfully, fairly and transparently (lawfulness, fairness and transparency).[66] Personal data that are collected for one purpose may not be used for incompatible purposes (purpose limitation).[67] Data must be adequate, relevant and limited to what is necessary in relation to the processing purposes (data minimisation).[68] Data must be 'accurate and, where necessary, kept up to date; every reasonable step must be taken to ensure that personal data that are inaccurate, having regard to the purposes for which they are processed, are erased or rectified without delay' (accuracy).[69] Data must be 'kept in a form which permits identification of data subjects for no longer than is necessary for the purposes for which

Yale Journal of Law and Technology 24 (2012); A Bradford, 'The Brussels Effect', Northwestern University Law Review (2012) 107 1.

[61] Article 1(1) of the Data Protection Directive. See also article 1(2) of the General Data Protection Regulation.

[62] Article 1(2) of the Data Protection Directive. See also article 1(3) of the General Data Protection Regulation.

[63] Article 2(a) of the Data Protection Directive. See also article 4(1) of the General Data Protection Regulation.

[64] Article 2(d) of the Data Protection Directive. See also article 4(7) of the General Data Protection Regulation.

[65] See on the Fair Information Practices: R Gellman, 'Fair Information Practices: A Basic History' (Version 2.16, June 17, 2016, continuously updated) <http://bobgellman.com/rg-docs/rg-FIPShistory.pdf> accessed 6 September 2017.

[66] Article 5(a) of the General Data Protection Regulation. Article 5 of the General Data Protection Regulation corresponds with Article 6 of the Data Protection Directive.

[67] Article 6(b) of the Data Protection Directive; Article 5(b) of the General Data Protection Regulation.

[68] Article 6(c) of the Data Protection Directive; Article 5(c) of the General Data Protection Regulation.

[69] Article 6(d) of the Data Protection Directive; Article 5(d) of the General Data Protection Regulation.

the personal data are processed' (storage limitation).[70] Appropriate security of personal data must be ensured (integrity and confidentiality).[71]

In 2018, the General Data Protection Regulation (GDPR) will replace the Data Protection Directive. While based on the same principles as the Directive, the Regulation brings significant changes. For instance, unlike a directive, a regulation has direct effect in the Member States. A regulation does not require implementation in the national laws of the Member States to be effective.[72] Hence, the Regulation should lead to a more harmonised regime in the European Union.[73] Moreover, the Regulation aims to improve compliance and enforcement. Under the Regulation, Data Protection Authorities can, in some situations, impose fines of up to 4% of a company's worldwide turnover.[74]

The Charter's right to protection of personal data and the right to privacy partly overlap. But in some respects, the right to protection of personal data has a broader scope than the right to privacy. The right to protection of personal data, and data protection law, apply as soon as personal data – any data relating to an identifiable person – are processed. Data protection law aims to ensure fairness when personal data are processed: such data must be processed 'lawfully, fairly and in a transparent manner in relation to the data subject'.[75] Data protection law deals with 'information privacy'[76] and 'data privacy',[77] but it also aims, for instance, to protect people against discriminatory effects of data processing.[78]

---

[70] Article 6(e) of the Data Protection Directive; Article 5(e) of the General Data Protection Regulation.
[71] Article 6(f) of the Data Protection Directive; Article 5(f) of the General Data Protection Regulation.
[72] Article 288 of the Treaty on the Functioning of the EU (consolidated version 2012).
[73] The extent to which the Regulation will actually harmonize data protection rules is a matter of discussion, see e.g. P. Blume, 'The Myths Pertaining to the Proposed General Data Protection Regulation', (2014) 4(4) International Data Privacy Law 269.
[74] Article 79 of the General Data Protection Regulation.
[75] Article 5(1)(a) of the General Data Protection Regulation, which replaces Article 6(1)(a) of the Data Protection Directive.
[76] DJ Solove & PM Schwartz, *Information Privacy Law* (Aspen 2014).
[77] LA Bygrave, *Data privacy law. An international perspective* (Oxford University Press 2014).
[78] See E Brouwer, *Digital Borders and Real Rights: Effective Remedies for Third- Country Nationals in the Schengen Information System (Martinus Nijhoff Publishers 2008)* 200. See also Recital 71 and Article 21 of the General Data Protection Regulation, on profiling and automated decisions. See about that provision the chapter by Boehm & Petkova in this book.

In some respects, privacy has a broader scope than data protection law. For example, a stalker often violates the victim's privacy. However, if the stalker does not collect or process the victim's personal data, data protection law does not apply.[79]

In conclusion, both the European Convention on Human Rights of the Council of Europe and the Charter of Fundamental Rights of the European Union protect privacy and freedom of expression.[80] The more recent Charter also explicitly protects the right to the protection of personal data.

## 4 The *Google Spain* judgment of the Court of Justice of the European Union

We now turn to the *Google Spain* judgment of the Court of Justice of the European Union to see how this court has applied the rights to privacy, data protection, and freedom of expression in a concrete case. The *Google Spain* judgment was triggered by a Spanish dispute between, on the one hand, Google, and on the other hand, Mr. Costeja González and the Spanish Data Protection Authority. Costeja González took issue with a link in Google's search results to a 1998 newspaper announcement concerning a real estate auction to recover his social security debts.[81] Without Google's search engine, the newspaper announcement would probably have faded from memory, hidden by practical obscurity.[82]

Mr. Costeja González wanted Google to delist the search result for searches on his name, because the information suggesting he had financial problems was outdated. Costeja González complained to the Spanish Data Protection Authority, which upheld the complaint against the search engine. Court proceedings commenced, and eventually a Spanish judge asked the Court of Justice of the European Union on guidance on how to interpret the Data Protection Directive.[83]

---

[79] See on the scope of privacy and data protection: G González Fuster, *The Emergence of Personal Data Protection as a Fundamental Right of the EU* (Springer 2014); FJ Zuiderveen Borgesius, *Improving Privacy Protection in the Area of Behavioural Targeting* (Kluwer Law International 2015), chapter 5, section 2; O Lynskey, *The foundations of EU data protection law* (Oxford University Press 2015).

[80] See: K Lenaerts and JA Gutiérrez Fons, 'The Place of the Charter in the EU Constitutional Edifice' in S Peers and others (eds), *The EU Charter of Fundamental Rights: A Commentary* (Hart Publishing 2014).

[81] See for the original publication: <http://hemeroteca.lavanguardia.com/preview/1998/01/19/pagina-23/33842001/pdf.html> accessed 6 September 2017.

[82] We borrow the 'practical obscurity' phrase from the US Supreme Court: *Dep't of Justice v. Reporters Comm. for Freedom of the Press*, 489 U.S. 749, 762 (1989). For an analysis see: KH Youm and A Park, 'The "Right to Be Forgotten" in European Union Law Data Protection Balanced with Free Speech?', (2016) 93(2) Journalism and Mass Communication quarterly 273.

In *Google Spain*, the Court of Justice of the European Union states that a search engine enables searchers to establish 'a more or less detailed profile' of a data subject, thereby 'significantly' affecting privacy and data protection rights.[84] According to the court, search results for a person's name provide 'a structured overview of the information relating to that individual that can be found on the internet – information which potentially concerns a vast number of aspects of his private life and which, without the search engine, could not have been interconnected or could have been only with great difficulty.'[85]

The Court of Justice of the European Union says that search engine operators process personal data if they index, store, and refer to personal data available on the web.[86] Moreover, the court sees search engine operators as 'data controllers' in respect of this processing.[87] Data controllers must comply with data protection law. The court also reaffirms that data protection law applies to personal data that are already public.

The Data Protection Directive contains provisions that aim to balance data protection interests and freedom of expression. For example, the directive provides for an exception for data that are processed for journalistic purposes or artistic and literary expression, if 'necessary to reconcile the right to privacy with the rules governing freedom of expression.'[88] But the court states that a search engine operator cannot rely on the exception in data protection law for data processing for journalistic purposes.[89]

---

[83] See on delisting requests in Spain: M Peguera, 'In the aftermath of *Google Spain*: how the 'right to be forgotten' is being shaped in Spain by courts and the Data Protection Authority' (2015) 23(4) International Journal of Law and Information Technology 325.
[84] Case C-131/12 *Google Spain v. Agencia Española de Protección de Datos (AEPD) and Mario Costeja González*, ECLI:EU:C:2014:317, paras 37 and 80.
[85] ibid, para 80.
[86] ibid, para 28.
[87] ibid, para 28.
[88] Article 9 Data Protection Directive. See also Article 85 of the General Data Protection Regulation.
[89] Case C-131/12 *Google Spain v. Agencia Española de Protección de Datos (AEPD) and Mario Costeja González*, ECLI:EU:C:2014:317, para 85. The English version of the judgment says Google 'does not appear' to be able to benefit from the media exception. However, in the authentic language of the judgment, Spanish, the CJEU says Google cannot benefit from the media exception. See on the media

The court holds that people have, under certain circumstances, the right to have search results for their name delisted. This right to have search results delisted also applies to lawfully published information. The court bases its judgment on the Data Protection Directive and the privacy and data protection rights of the Charter of Fundamental Rights of the European Union.[90] More specifically, the court bases its decision on the Data Protection Directive's provisions that grant data subjects, under certain conditions, the right to request erasure of personal data, and the right to object to processing personal data.[91]

The Data Protection Directive grants every data subject the right to correct or erase personal data that are not processed in conformity with the directive.[92] In *Google Spain*, the court clarifies that not only inaccurate data can lead to such unconformity, but also data that are 'inadequate, irrelevant or no longer relevant, or excessive' in relation to the processing purposes, for instance because the data have been stored longer than necessary.[93] In such cases, a search engine operator must delist the result at the request of the data subject.

The *Google Spain* judgment focuses on searches based on people's names. For instance, a search engine may have to delist an article announcing a public auction of a house at 10 Eye Street for a search for 'John Doe' who is mentioned in the article. But after a successful delisting request of John Doe, the search engine can still legally refer to the same article when somebody searches for '10 Eye Street'. Making a publication harder to find, but only for searches based on a name, reintroduces some practical obscurity:

---

exception: D Erdos, 'From the Scylla of Restriction to the Charybdis of Licence? Exploring the scope of the 'special purposes' freedom of expression shield in European data protection', (2015) 51(1) Common Market Law Review 119.

[90] Case C-131/12 *Google Spain v. Agencia Española de Protección de Datos (AEPD) and Mario Costeja González*, ECLI:EU:C:2014:317, para 99.

[91] Articles 12(b) and 14(a) of the Data Protection. See Articles 15, 17, and 21 of the General Data Protection Regulation.

[92] Article 12(b) and 14(a) of the Data Protection Directive. See also Article 16-19 of the General Data Protection Regulation.

[93] Case C-131/12 *Google Spain v. Agencia Española de Protección de Datos (AEPD) and Mario Costeja González*, ECLI:EU:C:2014:317, para 93.

the information is still available, but not as easily accessible in relation to the person's name.[94]

The Court of Justice of the European Union says in *Google Spain* that a 'fair balance' must be struck between the searchers' legitimate interests, and the data subject's privacy and data protection rights.[95] However, the court says that the data subject's privacy and data protection rights override, 'as a rule', the search engine operator's economic interests, and the public's interest in finding information.[96] With that 'rule', it seems that the Court of Justice of the European Union takes a different approach than the European Court of Human Rights, which says that freedom of expression and privacy have equal weight. However, the Court of Justice of the European Union makes this remark specifically in the context of delisting requests. Hence, the Court did not say that privacy and data protection rights generally override other rights.

The Court of Justice of the European Union also stresses that data subjects' rights should not prevail if the interference with their rights can be justified by the public's interest in accessing information, for example, because of the role played by the data subject in public life. This approach resembles the approach of the European Court of Human Rights when balancing privacy and the freedom of expression. As mentioned, the European Court of Human Rights considers how well-known the person is about whom a publication speaks. Public figures such as politicians must accept more interference with their privacy than ordinary citizens.

When a search engine operator delists a search result, freedom of expression may be interfered with in at least three ways.[97] First, those offering information, such as publishers and journalists, have a right to freedom of expression. As noted, the right to freedom of expression protects not only the expression (such as a publication), but also the means of communicating that expression.[98] Therefore, if the delisting makes it more difficult to find the publication, the freedom to impart information is interfered with.[99] Second, search engine users have a right to receive information. Third, a search engine operator exercises its freedom of expression when it presents its search results; an organised list of search results could be considered a form of expression.[100]

---

[94] See P Korenhof and L Gorzeman, 'Who Is Censoring Whom? An Enquiry into the Right to Be Forgotten and Censorship' July 15, 2015, <https://ssrn.com/abstract=2685105>.
[95] ibid, para 81.
[96] ibid, para 99.
[97] JVJ van Hoboken, *Search Engine Freedom. On the Implications of the Right to Freedom of Expression for the Legal Governance of Web Search Engines* (Kluwer Law International 2012) 350.
[98] *Autronic v Switzerland* App no 12726/87 (ECtHR 22 May 1990), para 47.

The *Google Spain* judgment was controversial. Many feared that freedom of expression would receive insufficient protection after the judgment. The NGO Index on Censorship said: 'The Court's decision (…) should send chills down the spine of everyone in the European Union who believes in the crucial importance of free expression and freedom of information.'[101] Others welcomed the judgment.[102]

In sum, the Court of Justice of the European Union recognised a right to be delisted. The right to be delisted requires from search engine operators that they delist, at the request of a data subject, outdated search results for name searches. But national courts and data protection authorities must decide on actual delisting requests. In the next section, we discuss how Google, Data Protection Authorities, and courts deal with delisting requests after the *Google Spain* judgment.

---

[99] JVJ van Hoboken, *Search Engine Freedom. On the Implications of the Right to Freedom of Expression for the Legal Governance of Web Search Engines* (Kluwer Law International 2012) 350.

[100] Case C-131/12 Google Spain v. Agencia Española de Protección de Datos (AEPD) and Mario Costeja González, Opinion of AG Jääskinen, para 132. JVJ van Hoboken, *Search Engine Freedom. On the Implications of the Right to Freedom of Expression for the Legal Governance of Web Search Engines* (Kluwer Law International 2012) 351. In the United States, some judges have granted search engines such freedom of expression claims (on the basis of the First Amendment of the US Constitution). E.g. *Search King, Inc. v. Google Technology, Inc.*, 2003 WL 21464568 (W.D. Okla. 2003). For a discussion see: E Volokh and DM Falk, 'Google First Amendment Protection for Search Engine Search Results' (2011-2012) 82 Journal of Law, Economics and Policy 883. For criticism on granting such claims, see: O Bracha, 'The Folklore of Informationalism: The Case of Search Engine Speech' (2014) 82 Fordham Law Review 1629.

[101] Index on Censorship, 'Index blasts EU court ruling on "right to be forgotten"' <https://www.indexoncensorship.org/2014/05/index-blasts-eu-court-ruling-right-forgotten> accessed 6 September 2017.

[102] See e.g. J. Powles, The Case That Won't Be Forgotten, Loyola University Chicago Law Journal, vol. 47, p. 583; H, Hijmans, *The European Union as Guardian of Internet Privacy: The Story of Art 16 TFEU*, Springer, 2016.

### 4.1 After the *Google Spain* judgment

### 4.2 Google

After the *Google Spain* judgment, Google created an online form that enables people to request the delisting of particular results for searches on their name.[103] If such a request is made, Google will 'balance the privacy rights of the individual with the public's interest to know and the right to distribute information.'[104] Google will look at 'whether the results include outdated information about you, as well as whether there's a public interest in the information — for example, we may decline to remove certain information about financial scams, professional malpractice, criminal convictions, or public conduct of government officials.'[105] Between fifty and one hundred people are working fulltime at Google to deal with delisting requests.[106]

As of August 2017, Google received over 588.000 requests and has evaluated more than 2.1 million URLs. Google has delisted roughly 43% of those URLs.[107] The top ten sites impacted by delisting requests include Facebook, YouTube, Twitter, and Profile Engine (a site that crawls Facebook).

Google gives 23 examples of how it dealt with delisting requests. Examples of granted requests, quoted from Google, include:

- 'An individual who was convicted of a serious crime in the last five years but whose conviction was quashed on appeal asked us to remove an article about the incident.'

---

[103] Google, 'Removing content from Google' <https://support.google.com/legal/troubleshooter/1114905?hl=en#ts=1115655%2C6034194> accessed 6 September 2017.

[104] Google, 'Search removal request under data protection law in Europe' <https://support.google.com/legal/contact/lr_eudpa?product=websearch> accessed 6 September 2017.

[105] ibid.

[106] As reported by Peter Fleischer, Google's Global Privacy Counsel, at the Privacy & Innovation Conference at Hong Kong University, 8 June 2015, <www.lawtech.hk/pni/?page_id=11> accessed 6 September 2017.

[107] Google, 'European privacy requests for search removals' <www.google.com/transparencyreport/removals/europeprivacy/?hl=en> accessed 6 September 2017.

- 'A woman requested that we remove pages from search results showing her address.'
- 'A victim of rape asked us to remove a link to a newspaper article about the crime.'
- 'A man asked that we remove a link to a news summary of a local magistrate's decisions that included the man's guilty verdict. Under the UK Rehabilitation of Offenders Act, this conviction has been spent.'[108]

In all these cases Google delisted the search result for the individual's name.

Examples of denied requests include:

- 'We received a request from a former clergyman to remove 2 links to articles covering an investigation of sexual abuse accusations while in his professional capacity.'
- 'An individual asked us to remove a link to a copy of an official state document published by a state authority reporting on the acts of fraud committed by the individual.'
- 'An individual asked us to remove links to articles on the internet that reference his dismissal for sexual crimes committed on the job.'[109]

In all these cases, Google denied the request.

The examples suggest that Google does a reasonable job when dealing with delisting requests. However, Google could be more transparent about how it deals with delisting requests. More transparency could enable regulators, academics, and members of the public to keep an eye on removal practices. As noted, Google delisted over 785.000 URLs. It is unclear whether those URLs concerned news articles, blog posts, revenge porn, or other materials. Moreover, we do not know whether requests mainly come from ordinary citizens, politicians, criminals, or other people.

---

[108] ibid.
[109] ibid.

## 4.3 Data Protection Authorities

The EU's national Data Protection Authorities cooperate in the Article 29 Working Party, an advisory body.[110] The Working Party published guidelines on the implementation of the *Google Spain* judgment. The Working Party says that '[i]n practice, the impact of the de-listing on individuals' rights to freedom of expression and access to information will prove to be very limited.'[111] Nevertheless, Data Protection Authorities 'will systematically take into account the interest of the public in having access to the information.'[112] The Working Party also called on search engine operators to be transparent about their decisions: 'the Working Party strongly encourages the search engines to publish their own de-listing criteria, and make more detailed statistics available.[113]

The Working Party developed a set of criteria to help Data Protection Authorities to assess, on a case-by-case basis, whether a search engine operator should delist a search result. The Working Party states that '[i]t is not possible to establish with certainty the type of role in public life an individual must have to justify public access to information about them via a search result.'[114] Nevertheless, the Working Party says that 'politicians, senior public officials, business-people and members of the (regulated) professions' can usually be considered to play a role in public life. Regarding minors, the Working Party notes that Data Protection Authorities are more inclined to delist results.[115] Data Protection Authorities are also more likely to intervene if search results reveal 'sensitive data.'[116] The criteria developed by the Working Party offer guidance to both search engines and Data Protection Authorities when they decide on de-listing requests.[117]

---

[110] See S Gutwirth and Y Poullet, 'The contribution of the Article 29 Working Party to the construction of a harmonised European data protection system: an illustration of 'reflexive governance'?' in VP Asinari P and Palazzi (eds), *Défis du Droit à la Protection de la Vie Privée. Challenges of Privacy and Data Protection Law* (Bruylant 2008).
[111] Article 29 Working party, 14/EN WP 225 (2014) 2 and 6.
[112] ibid, 2.
[113] ibid, 3 and 10.
[114] ibid, 13.
[115] ibid, 15.
[116] See on sensitive data: Section 5.3.2.
[117] See for an analysis of factors to take into account when deciding on delisting requests: J Ausloos and A Kuczerawy, 'From Notice-and-Takedown to Notice-and-Delist: Implementing the Google Spain Ruling' (2016) 14(2) Colorado Technology Law Journal 219.

## 4.4 Open questions

Below we discuss some open questions regarding delisting requests after *Google Spain*. More specifically, we consider how the right to be forgotten relates to public registers and open data policies. We also highlight the uncertainties surrounding sensitive personal data the processing of such data by search engines. And we consider whether search engines should delist search results only on EU-domains, or also on other domains.

### *4.4.1 Public registers and open data*

Nowadays, many personal data are made accessible online. For instance, through open data initiatives and public registers, personal data are generally available on the web. If a public register is published online, its data can be collected and republished by data brokers, journalists, search engines, and others. Such data re-use can serve important goals, such as fostering transparency, innovation, and public sector efficiency. However, data re-use can also threaten privacy.[118] Questions also rise about how the publication of such data relates to right to be forgotten. Information may have to be removed, depending on factors such as the type of information, the information's relevance, and the time that has elapsed.

The Court of Justice of the European Union has given the beginning of an answer in the *Manni* case.[119] This case dealt with the removal of information in the official chamber of commerce of Lecce in Italy. Mr. Manni asked the chamber of commerce to erase, anonymise, or block the information linking him to the liquidation of a company in 1992. According to Manni, he cannot sell properties of a tourist complex, because potential purchases have access to that information in the company register.

In this case, the Italian Court of Cassation asked the Court of Justice of the European Union how the principle of data minimisation related to the duty for Member States to disclose information in a company register under the Company Law Directive.[120] The Italian court also asked whether Member States may limit access to information about dissolved companies. To determine whether people have a right to erasure, the Court of Justice of the European Union considers the purpose of the companies register. The disclosure of information as prescribed by the Company Law Directive essentially aims to protect the interests of third parties in relation to companies with limited liability.[121] The directive is silent on whether information should also be available after a company has dissolved. Yet, according to the court, information must remain available because rights and legal relations may continue to exist.[122] Moreover, given the diversity of national law regarding time limits, the court finds that it is impossible to identify a certain period of time that must elapse before someone has the right to obtain erase information or the blocking access to it.[123]

---

[118] FJ Zuiderveen Borgesius, J Gray and M van Eechoud, 'Open Data, Privacy, and Fair Information Principles: Towards a Balancing Framework' (2015) 30 Berkeley Technology Law Journal 2073.
[119] Case C-131/12 *Camera di Commercio, Industria, Artigianato e Agricoltura di Lecce v. Salvatore Manni*, ECLI:EU:C:2017:197.

The Court of Justice of the European Union also considers the right to privacy and the right to protection of personal data, as protected by the Charter of Fundamental Rights of the European Union.[124] The court finds that these rights are not disproportionately interfered with for two reasons. First, because the types of information are limited to information about a particular person in relation to the company.[125] Second, because those who trade with joint-stock companies and limited liability companies have as safeguards only the assets of the latter companies. The court notes that 'it appears justified that natural persons who choose to participate in trade through such a company are required to disclose the data relating to their identity and functions within that company, especially since they are aware of that requirement when they decide to engage in such activity.'[126]

[120] First Council Directive 68/151/EEC of 9 March 1968 on co-ordination of safeguards which, for the protection of the interests of members and others, are required by Member States of companies within the meaning of the second paragraph of Article 58 of the Treaty, with a view to making such safeguards equivalent throughout the Community (OJ 1968 L 65, p. 8), as amended by Directive 2003/58/EC of the European Parliament and of the Council of 15 July 2003 (OJ 2003 L 221, p. 13).
[121] Case C-131/12 *Camera di Commercio, Industria, Artigianato e Agricoltura di Lecce v. Salvatore Manni*, ECLI:EU:C:2017:197, paras 49 and 51.
[122] ibid, para 53.
[123] ibid, paras 54-56.
[124] Art. 7 and 8 of the Charter of Fundamental Rights of the European Union.
[125] Case C-131/12 *Camera di Commercio, Industria, Artigianato e Agricoltura di Lecce v. Salvatore Manni*, ECLI:EU:C:2017:197, para 58.

Yet, the Court of Justice of the European Union also notes that 'it cannot be excluded' that in exceptional cases, access to information is limited to parties who have a specific interest in the information, but only after a 'sufficiently long period' after the dissolution of a company.[127] In such cases, people may have a right to object against the processing of their data.[128] But the court also stresses that Member State law may restrict this right to object.[129] According to the Court of Justice of the European Union, the Italian court must determine the extent to which Italian law enables or limits the right to object, and whether the facts and circumstances of the case justify the limitation of access to the data concerning Manni.[130] The Court of Justice of the European Union also highlights that 'the mere fact that (…) the properties of a tourist complex built by Italiana Costruzioni, of which Mr Manni is currently the sole director, do not sell because of the fact that potential purchasers of those properties have access to that data in the company register' does not constitute a reason to limit to the information in the companies register.[131]

In sum, the Court of Justice of the European Union requires, in principle, that the personal data must be available in the companies register, even after a company has dissolved. Yet, the court also leaves room for Member States to enable individuals to apply, in exceptional cases, for a limitation of the availability of information about them in company registers.

### *4.4.2 Sensitive data*

The *Google Spain* judgment has caused a problem regarding search engine operators and 'special categories of data'. Such special categories of data are 'personal data revealing racial or ethnic origin, political opinions, religious or philosophical beliefs, trade-union membership, and the processing of data concerning health or sex life.'[132] Regarding data relating to offences and criminal convictions, the Data Protection Directive states that processing 'may be carried out only under the control of official authority, or if suitable specific safeguards are provided under national law, subject to derogations which may be granted by the Member State under national provisions providing suitable specific safeguards.'[133] All these categories of data receive extra protection in the Data Protection Directive, because such data 'are capable by their nature of infringing fundamental freedoms or privacy'.[134] For brevity, we refer to 'sensitive data', rather than to special categories of data.

---

[126] ibid, para 59.
[127] ibid, para 60.
[128] Art. 14(a) of the Data Protection Directive.
[129] Case C-131/12 *Camera di Commercio, Industria, Artigianato e Agricoltura di Lecce v. Salvatore Manni*, ECLI:EU:C:2017:197, paras 61 and 62.
[130] ibid, para 62-63.
[131] ibid, para 63.

As noted, the Court of Justice of the European Union chose to see search engines operator as data controllers when they index, store, and refer to personal data on websites. That choice has caused a problem with sensitive data. The Data Protection Directive only allows personal data processing if the controller can rely on a legal basis for processing.[135] In *Google Spain*, the Court of Justice of the European Union ruled that, for the processing at issue, a search engine could rely on the legitimate interests provision.[136] This provision, also called the balancing provision, permits processing if the controller's legitimate interests, or those of a third party, outweigh the data subject's fundamental rights.

However, the processing of sensitive data is only allowed after the data subject gave his or her explicit consent, or if an exception applies that can legalise the processing of sensitive data.[137]

---

[132] Art. 8(1) of the Data Protection Directive. See also article 9(1) of the General Data Protection Regulation.
[133] Article 8(5) of the Data Protection Directive. See also article 10 of the General Data Protection Regulation.
[134] Recital 33 of the Data Protection Directive.
[135] Article 8(2) of the Charter of Fundamental Rights of the European Union; Article 7 of the Data Protection Directive; article 6 of the General Data Protection Regulation.
[136] Article 7(f) of the Data Protection Directive. Case C-131/12 *Google Spain v. Agencia Española de Protección de Datos (AEPD) and Mario Costeja González*, ECLI:EU:C:2014:317, para 73.
[137] Article 8 of the Data Protection Directive. See also article 9 of the General Data Protection Regulation. In some member states, data subjects cannot override the in-principle prohibition of processing sensitive data by giving their explicit consent. See European Commission, Commission Staff Working Paper, Impact Assessment Accompanying the proposal for the General Data Protection Regulation, SEC(2012)

In practice, many websites that are indexed by a search engine may include sensitive data. For example, a website might include a picture of an identifiable person in a wheelchair: personal data concerning that person's health. And the website of a Catholic choir may include a member list: personal data indicating religion.

Because processing sensitive data requires explicit consent, a search engine operator would need the data subject's explicit consent for processing web pages that include such data. Asking hundreds of millions of people for their consent would be impossible. The Data Protection Directive contains exceptions to the in-principle prohibition of processing sensitive data, such as exceptions for churches and for the medical sector. But in many cases, search engine operators cannot rely on any of the exceptions.[138]

Therefore, it seems that a search engine operator's practices are, formally, partly illegal – when the operator processes sensitive data included on web pages, and no exceptions apply. That formal illegality is a side-effect of the choice of the Court of Justice of the European Union to see a search engine operator as a controller regarding the processing of personal data on third party web pages.

To avoid that search engine activities would be rendered partly illegal, the Advocate General in *Google Spain* did not want to regard a search engine operator as a controller regarding the processing of personal data made available on third party web pages.[139] In contrast, the Court of Justice of the European Union did not explicitly consider consequences of seeing search engine operators as data controllers. After the *Google Spain* judgment, Google, Data Protection Authorities, and courts have solved the sensitive data problem by ignoring it.

But a 2016 judgment by a Dutch lower court (which was later overruled) illustrates how this sensitive data problem may play out. An attorney submitted a delisting request to Google, regarding a blog post about a criminal conviction of the attorney in another country. Under Dutch law, personal data regarding criminal convictions are sensitive data.[140] The court held that Google could not link to the blog post, because the post contained sensitive data. Google had not asked the attorney for his consent for referring to the blog post, and therefore Google could not legally link to it.[141] Google appealed the decision.

---

72 final, Annex 2, <http://ec.europa.eu/justice/data-protection/document/review2012/sec_2012_72_en.pdf> accessed 6 September 2017, p. 29.

[138] See in more detail: S Kulk and FJ Zuiderveen Borgesius, 'Google Spain v. Gonzalez: Did the Court Forget about Freedom of Expression' (2014) 5(3) European Journal of Risk Regulation 389.

[139] Case C-131/12 Google Spain v. Agencia Española de Protección de Datos (AEPD) and Mario Costeja González, Opinion of AG Jääskinen, para 90.

The Dutch Court of Appeals mitigated the sensitive data problem. In short, the Court of Appeals decided that Google can rely on an exception for journalistic purposes. (The Court of Appeals seems to come to a decision that is contrary to the Court of Justice of the European Union.[142]) Hence, the Court of Appeals says that, under certain conditions, Google is allowed to process sensitive data for its search engine.[143]

Still, a coherent Europe Union-wide solution must be found for the sensitive data problem. Perhaps the Court of Justice of the European Union can find such a solution. A French judge asked preliminary questions to the Court of Justice of the European Union, regarding this sensitive data problem in the context of delisting requests. Perhaps the Court of Justice of the European Union can think of a creative solution to solve the problem.[144]

Otherwise, the EU legislator must take action. One possibility would be to introduce a new exception to the in-principle prohibition of processing sensitive data, specifically for search engines. However, it does not seem plausible that the EU lawmaker will revise the General Data Protection Regulation anytime soon.

---

[140] Article 16 of the Dutch Data Protection Act.

[141] Rechtbank Rotterdam, 29 March 2016, ECLI:NL:RBROT:2016:2395 <http://deeplink.rechtspraak.nl/uitspraak?id=ECLI:NL:RBROT:2016:2395>. See FJ Zuiderveen Borgesius, 'Het 'right to be forgotten" en bijzondere persoonsgegevens: geen ruimte meer voor een belangenafweging?' ['The 'right to be forgotten' and sensitive personal data: no room for balancing?'], Computerrecht 2016-4, 220.

[142] See section 4 of this chapter.

[143] Hof Den Haag, 23 May 2017, www.rechtspraak.nl, ECLI:NL:GHDHA:2017:1360.

[144] See <http://english.conseil-etat.fr/Activities/Press-releases/Right-to-be-delisted> accessed 6 September 2017.

### *4.4.3 Delisting requests only for EU domains?*

The domain-related scope of the right to be delisted is contentious.[145] Google chose to delist search results only on its European domains (e.g. google.de or google.fr). Hence, Google did not delist search results on its google.com domain.[146] The Article 29 Working Party, however, says that 'limiting de-listing to EU domains on the grounds that users tend to access search engines via their national domains cannot be considered a sufficient mean to satisfactorily guarantee the rights of data subjects according to the [*Google Spain*] ruling.'[147]

It is difficult to defend that the domain name of a search engine website should be the main factor when deciding which national law applies. If that were the main factor, a search engine operator could easily escape the application of national laws by opting for a particular domain name.[148] However, Google – a company from the US, where freedom of expression is strongly protected – does not want to delist search results on its .com domain. Google argues that it usually sends users from Europe to their local domain, for instance Google.fr for France and Google.de for Germany.[149]

The Court of Justice of the European Union will have to decide about the domain-related scope of delisting requests. In 2015 the French Data Protection Authority, *Commission Nationale de l'Informatique et des Libertés* (CNIL) demanded that Google delisted results not only for searches on their European domains such as Google.fr, but on all its domains, including google.com. According to CNIL, 'to be effective, delisting must be carried out on all extensions of the search engine.'[150]

---

[145] B Van Alsenoy and M Koekkoek, 'Internet and jurisdiction after Google Spain: the extraterritorial reach of the 'right to be delisted' 5(2) International Data Privacy Law 105; C Kuner, 'The Court of Justice of the EU judgment on data protection and internet search engines: current issues and future challenges' in: H Hijmans and H Kranenborg (eds.), *Data protection anno 2014: how to restore trust?* (Intersentia 2014).
[146] See: J Powles, 'The case that won't be forgotten', 47 Loyola University Chicago Law Journal 583, 596.
[147] Article 29 Working party, 14/EN WP 225 (2014) 3.
[148] Y Fouad, 'Reikwijdte van het Europese dataprotectierecht na Google Spanje: wat is de territoriale werkingssfeer en wordt eenieder beschermd?' ['Scope of the European data protection law after *Google Spain*: what is the territorial scope and is everyone protected?'], Master thesis 2015, Institute for Information Law (University of Amsterdam), on file with authors.
[149] P Fleischer, 'Reflecting on the Right to be Forgotten' Google In Europe Blog <www.blog.google/topics/google-europe/reflecting-right-be-forgotten> accessed 6 September 2017.

Google did not comply. CNIL responded by starting formal proceedings against Google. In the meantime, Google implemented geo-location technology to ensure search results are delisted on Google domains if these domains are accessed from the country of the data subject.[151] To illustrate: an internet user with a French IP address would not see delisted search results on Google.com. An internet user with an IP address from the US would see all search results on Google.com, including results that were delisted in France.

CNIL, however, was not satisfied, as the delisted results would still be available to search engine users outside France. Moreover, people could circumvent the geo-block by using a virtual private network, which enables people to access a website using a foreign IP-address.[152] CNIL fined Google 100.000 Euros. Google appealed the decision with the French *Conseil d'État* – the highest administrative court in France.[153] The *Conseil d'État* has asked the Court of Justice of the European Union preliminary questions. In short, the Court of Justice of the European Union is asked for the correct interpretation of data protection law regarding the domain-related scope of delisting requests.[154]

---

[150] CNIL, 'CNIL orders Google to apply delisting on all domain names of the search engine' <www.cnil.fr/en/cnil-orders-google-apply-delisting-all-domain-names-search-engine> accessed 6 September 2017.

[151] Google, 'Adapting our approach to the European right to be forgotten' <https://europe.googleblog.com/2016/03/adapting-our-approach-to-european-right.html> accessed 6 September 2017.

[152] CNIL, decision no. 2016-054, unofficial translation by CNIL <www.cnil.fr/sites/default/files/atoms/files/d2016-054_penalty_google.pdf> accessed 6 September 2017.

[153] M Scott, 'Google Appeals French Privacy Ruling' *New York Times* (New York, 19 May 2016) <www.nytimes.com/2016/05/20/technology/google-appeals-french-privacy-ruling.html> accessed 6 September 2017.

[154] Le Conseil d'État, 'Portée territoriale du droit au déréférencement' <http://www.conseil-etat.fr/Actualites/Communiques/Portee-territoriale-du-droit-au-dereferencement> accessed 6 September 2017.

## 5 Concluding thoughts

The *Google Spain* judgment of the Court of Justice of the European Union illustrates how, under European law, freedom of expression can be limited by privacy and data protection rights. Privacy and data protection rights require that, under certain conditions, search engine operators must delist outdated search results for name searches, if the relevant individual requests delisting. Yet, when assessing whether a search result must be delisted, search engine operators, Data Protection Authorities and national courts must also consider the extent to which delisting affects freedom of expression. Searchers have the right to receive information, and publishers, bloggers, journalists etc. have the right to impart information. National courts and Data Protection Authorities must therefore consider all relevant facts and circumstances and decide on a case-by-case basis whether a particular search result should be delisted for name searches.

In principle, privacy and freedom of expression have equal weight in Europe – which right prevails depends on the circumstances of a case. Balancing privacy-related and freedom of expression-related interests will always remain contentious and difficult. A case-by-case analysis is required, and all circumstances of a case should be taken into account. The Court of Justice of the European Union gave limited guidance as to when a search result should be delisted. But, the European Court of Human Rights has developed rich and nuanced case law on balancing privacy and freedom of expression, in which it provides a list of criteria to determine the right balance. We can expect much more case law on delisting requests; hopefully that will give more guidance for deciding about delisting requests.

The right to be forgotten discussion is part of a broader discussion on how to balance different rights and interests when personal information is accessible online. Similar questions arise with regard to, for instance, open data initiatives, and online archives of newspapers or case law. The discussion on the right to be forgotten is only the beginning of how the right to privacy and freedom of expression should be weighed in the online world.

* * *